\documentclass{article}
\usepackage{PRIMEarxiv}
\usepackage{amsmath}
\usepackage[utf8]{inputenc} 
\usepackage[T1]{fontenc}    
\usepackage{hyperref}       
\usepackage{url}            
\usepackage{booktabs}       
\usepackage{amsfonts}       
\usepackage{nicefrac}       
\usepackage{microtype}      
\usepackage{natbib}
\usepackage{lipsum}
\usepackage{fancyhdr}       
\usepackage{graphicx}       
\graphicspath{{media/}}     
\usepackage{xcolor}
\hypersetup{
    colorlinks,
    linkcolor={red!0!black},
    citecolor={blue!100!black},
    urlcolor={blue!0!black}
}

\title{Learning Nudges for Conditional Cooperation: A Multi-Agent Reinforcement Learning Model
}

\author{
  Shatayu Kulkarni, Sabine Brunswicker \\
  \href{www.rcodi.org}{Research Center for Open Digital Innovation (RCODI)} \\
  Purdue University \\
  West Lafayette, IN, USA\\
  \texttt{shatayu@alumni.purdue.edu, sbrunswi@purdue.edu} \\
}

\begin{document}
\maketitle

\begin{abstract}
The public goods game describes a social dilemma in which a large proportion of agents act as conditional cooperators (CC): they only act cooperatively if they see others acting cooperatively because they satisfice with the social norm to be in line with "what others are doing" instead of optimizing cooperation. CCs are guided by aspiration-based reinforcement learning guided by past experiences of interactions with others and satisficing aspirations. In many real-world settings, reinforcing social norms do not emerge in the first place, causing defect to take hold. In this paper, we put forward the argument that an optimizing reinforcement agent can facilitate cooperation by acting as a "social planner" using "nudges", i.e. indirect mechanisms for cooperation to happen. The agent's goal is to motivate CCs into cooperation through its own actions in order to create social norms that signal that others are cooperating. We propose a multi-agent reinforcement learning model for public goods games, with 3 CC learning agents using aspirational reinforcement learning and 1 nudging agent who uses deep reinforcement learning to learn "nudges" that optimizes cooperation in the public goods game. For our nudging agent, we model two distinct reward functions, one maximizing the total game return (sum DRL) and one maximizing the number of "cooperative contributions" - contributions higher than a proportional threshold (prop DRL). Our results show that our aspiration-based RL model for CC agents is consistent with empirically observed CC behavior. Furthermore, games combining 3 CC RL agents and one nudging RL agent outperform the baseline consisting of 4 CC RL agents only. The sum DRL nudging agent increases the total sum of contributions by 8.22\% and the total proportion of cooperative contributions by 12.42\%, while the prop nudging DRL increases the total sum of contributions by 8.85\% and the total proportion of cooperative contributions by 14.87\%. Our findings advance the literature on public goods games and multi-agent reinforcement learning with mixed incentives.
\end{abstract}



\section{Introduction}
Resolving intertemporal social dilemmas of collaboration is a central challenge in modern society, as social dilemmas create a trade-off between maximizing short-term individual gains and long-term collective benefits \citep{ostrom_elinor_conversation_2000, hardin_tragedy_1968}. Intertemporal public goods games, widely studied in the fields of computational economics, biology, social science, and also machine learning, are simplified, abstracted versions of intertemporal social dilemmas of multi-agent cooperation, in which players iteratively contribute individual assets (e.g., money, tokens, etc.) to a common pool, from which each player benefits equally at the end of the multi-round game \citep{lang_explaining_2018, ackermann_explaining_2019}. Formal game-theoretical analysis predicts that public goods agents—natural and artificial ones—act individually and fail to cooperate due to the optimization of short-term gains over long-term utility \citep{hardin_tragedy_1968}. However, empirical studies suggest that this does not always hold true. In some cases, even though they are rare, the game’s agents may actually find models to cooperate, e.g., when finding ways to sustainably use a common local fishery instead of overfishing \citep{jentoft_working_2018}, organizing collective food storage in preparation for a harsh winter \citep{hughes_inequity_2018}, or succeeding in collaborative open source software development. Such studies suggest that the success of collaboration in public goods games depends on the behavior of so-called conditional cooperators (CC): CC behavior "depends on the subjects' perception of future interaction" of other group members, and such perceptions relate to so-called "social norms" \citep{keser_conditional_2000, te_velde_conformity_2022, cialdini_descriptive_2007}. Social norms cast a conforming effect on players subject to them \citep{te_velde_conformity_2022, cialdini_descriptive_2007}. In public goods games, descriptive social norms quickly emerge from the observed average contribution of the other group members in the previous period, changing the game environment of the agents, causing positive or negative game dynamics in terms of cooperation \citep{keser_conditional_2000}. Typically, human CCs, or in a more general way "natural" CC agents, would adjust their individual aspirations to cooperate depending on the average contributions of the other players of the last round, following principles of aspiration-based reinforcement learning rooted in statistical learning theories developed in psychology and economic game theory \citep{mosteller_stochastic_1957, sutton_reinforcement_2020}. Emerging social norms often fail to create the positive game dynamics needed to increase cooperation \citep{ezaki_reinforcement_2016}. However, if there were a way that a social planner could indirectly modulate the emerging social norms (e.g., through its own actions), a mechanism also called \emph{nudging}, such actions could incentivize natural CC agents to cooperate \citep{andi_nudging_2021}.

Typically, human CCs, or in a more general way "natural" CC agents, would adjust their individual aspirations to cooperate depending on the average contributions of the other players of the last round, following principles of aspiration-based reinforcement learning rooted in statistical learning theories developed in psychology and economic game theory \citep{mosteller_stochastic_1957, sutton_reinforcement_2020}. Aspirational RL agents are not seeking to optimize utility but instead learn based on past experiences while being guided by aspirations shaped by social norms emerging bottom-up from the contribution behavior of all agents.

Existing work on machine learning has made significant progress in reinforcement learning (RL) algorithms that achieve cooperation in multi-agent reinforcement learning (MARL) among self-interested agents, who cannot be coordinated with a centralized RL algorithm \citep{canese_multi-agent_2021, hughes_inequity_2018}. Finding ways for self-interested agents to incentivize each other has shown promising results for solving the problem of cooperation among self-interested agents in MARL \citep{yang_learning_2020,munoz_de_cote_learning_2006,hughes_inequity_2018}.

There are also recent efforts to design a social planner who leverages deep graph-based reinforcement learning to nudge human-like agents into cooperation by modulating the dyadic information exchange among agents in a social network \citep{mckee_scaffolding_2023}. However, current research has not investigated the specific problem of how to design an artificial deep RL agent that acts as a social planner who seeks to optimize cooperation through its own contribution behavior to "nudge" human CCs into cooperation without formal intervention and in the absence of dyadic information exchange among CC agents. So the question that remains unanswered is: \emph{How can an artificial RL agent nudge natural CC agents into cooperative behavior to support cooperation while mitigating the risk that defection will take hold?} 

To tackle this gap, this paper studies multi-agent reinforcement learning games (MARL) for intertemporal public goods combining multiple aspirational CC agents and one deep reinforcement learning agent (DRL) who seeks to optimize cooperation. The DRL agent learns a policy for nudging the CC agents into cooperation through its own contributions. We model the learning of aspirational CC agents using aspirational reinforcement learning to emulate human CC behavior. For our nudging DRL agent, we design two reward functions: one seeking to optimize the total contribution to the game (sum DRL), and one optimizing the percentage of cooperative contributions in each round (prop DRL). We train our DRL agent using the Proximal Policy Optimization algorithm (PPO). In our experiments, we compare baseline games with CC RL agents only to mixed MARL games with multiple CC RL agents and one nudging DRL agent. We present four major findings:
\begin{enumerate}
    \item Our aspiration-based RL model for CC agents is consistent with empirically observed CC behavior of humans.
    \item For our social planner, our sum DRL agent slightly outperforms the prop DRL agent during training (in terms of convergence, loss, and entropy).
    \item In our games with 3 CC RL agents and one nudging RL agent, the sum DRL nudging agent increases the total sum of contributions by 8.22\% and the total proportion of cooperative contributions by 12.42\% compared to the baseline.
    \item In our games with 3 CC RL agents and one nudging RL agent, the prop DRL nudging agent increases the total sum of contributions by 8.85\% and the total proportion of cooperative contributions by 14.87\%, compared to the baseline.
\end{enumerate}

\section{Related Work} \label{sec:related work}

Our research relates to the work on 1) multi-agent reinforcement learning (MARL) and cooperation and 2) aspiration-based reinforcement learning in evolutionary game theory.

\subsection{Multi-agent Reinforcement Learning and Cooperation}
Achieving cooperation in multi-agent reinforcement learning remains a difficult problem \citep{canese_multi-agent_2021, jaques_social_2019, gronauer_multi-agent_2022, zhang_multi-agent_2021}. Prior work resorts to centralized learning strategies for decentralized policies have shown to improve cooperative game outcomes \citep{canese_multi-agent_2021}. However, such an approach is unrealistic in intertemporal social dilemma games as such games lack central coordination among autonomous agents. Recent MARL research has investigated new learning approaches to improve cooperation without centralized learning. For example, machine learning scholars have recently drawn on the theory of social psychology and designed social learning algorithms, in which agents use counterfactual reasoning to assess whether their actions will increase the chances of cooperation \citep{jaques_social_2019}. \citet{tampuu_multiagent_2017}, for example, designed a deep-Q-network game in which two autonomous Deep Q-Learning agents learn to cooperate and compete by sharing a high-dimensional environment and being only fed with raw visual input (e.g., screen images in a video game). They show that successful strategies for competition and cooperation emerge, depending on the incentives provided by the reward scheme. Research on MARL also specifically examines different algorithms for policy learning to tackle the problem of non-stationary environments. Indeed, finding efficient algorithms for deep RL methods is challenging.

MARL research also specifically focuses on social dilemmas: \citet{munoz_de_cote_learning_2006}, for example, propose a Q-learning algorithm for social dilemma games, in which self-interested Q-Learning agents motivate each other using "change or learn fast" and "change and keep" strategies to reduce learning rates. Both strategies end up encouraging more cooperation than unadjusted Q-Learning on its own. \citet{hughes_inequity_2018} design an MARL for intertemporal social dilemmas using an asynchronous advantage actor-critic (A3C) as the deep RL algorithm to study how agents with inequity-averse preferences impact cooperation among the agents. Some recent work also examines specifically social network relationships and dyadic communication between agents in social dilemma games. For example, \citet{mckee_scaffolding_2023} designs a social planner who leverages deep graph-based reinforcement learning to break or recommend connections between agents in an MARL game with social network relationships along which information is shared. However, the problem of designing a "social planner" leveraging deep RL to nudge cooperation among CCs using descriptive social norms in the absence of dyadic information exchange has yet to be explored. This paper tackles this gap.

\subsection{Aspiration-based Reinforcement Learning and Economic Game Theory}
Aspiration-based reinforcement learning has its roots in statistical learning theory in psychology and economics \citep{mosteller_stochastic_1957, sutton_reinforcement_2020, cross_stochastic_1973}. Inspired by statistical learning theory in psychology, economists started to adopt Herbert Simon's theory of "satisficing" \citep{simon_behavioral_1955, simon_decision_1987, march_variable_1988} arguing that humans are boundedly rational: They engage in reinforcement learning according to which humans satisfy rather than maximize payoffs in the sense of economic rationality \citep{bendor_aspiration-based_2001}. According to this principle, humans make choices based on past experiences and interactions with their decision environment that can be represented by probabilistic rules: Actions that have resulted in satisfactory payoffs are more likely to be selected compared to those that lead to unsatisfactory ones. To make such "judgments", humans are guided by aspirations that are socially defined (e.g., by the social norms and information available about others). Empirical studies have provided vast evidence that natural agents use aspiration-based reinforcement learning guided by simple heuristics instead of utility maximization, in particular in environments where it is almost impossible to form a coherent model of the environment. One model that shaped the field of aspiration-based reinforcement learning in the field of economic game theory was the Bush-Mosteller model \citep{mosteller_stochastic_1957}: The Bush-Mosteller model assumes that if an agent's action is followed by a positive payoff, the probability that the action is taken increases based on simple linear probabilistic rules of learning.

Research on computational economics and game theory has studied aspiration-based reinforcement learning in dynamic multi-agent games with the goal to facilitate cooperation. \citet{zhang_multi-agent_2021}, for example, studied aspiration-based RL and the Fermi-function-based strategy adoption rule in the context of rhe Prisoner's Dilemma and Snowdrift games. \citet{stahl_aspiration-based_2002} examined how aspiration-based RL and reciprocity-based RL influence the dynamics in MARL games to empirically explain cooperation behavior. \citet{song_reinforcement_2022}, for example, use the classical BM model in evolutionary game theory to study the effect of adaptive interaction intensity allowing direct information exchange between players. They find that an increased interaction intensity increases cooperation because it shapes the co-evolutionary process and also the microscopic mechanisms of cooperation. Aspiration-based RL has also been studied in the context of spatial public goods games \citep{tomassini_computational_2021}. Recently, research has also started to use aspiration-based RL to model CC behavior, and also its "moody" cousin. \citet{ezaki_reinforcement_2016}, for example, used a refined version of the classical RL model, the Bush-Mosteller (BM) model, to model CC behavior as aspiration-based RL, in which the agents' actions are constituted by the average contribution of the players in the last round, or in other words by social norms. Thus, they suggest social norms play an important role in modulating CCs' actions. However, as of today, there has been little effort to investigate how a social "planner", modeled as a deep RL agent, seeks to optimize cooperation by shaping the behavior of CC agents through its contribution. This paper seeks to fill this gap.

\section{The Public Goods Game and the Reinforcement Learning Agents}\label{sec2}

In the following sections, we first introduce the public goods game. Afterwards, we specify the reinforcement learning models of our human-like CC RL agents, and the social planner, the nudging deep RL agent. 

\subsection{Game Setup}

Following the structure used in seminal works \citep{ezaki_reinforcement_2016, horita_reinforcement_2017, fischbacher_are_2001}, our game involves \(N\) players (agents) participating in a total of \(t_{\text{max}}\) rounds. In our game, we chose \(N=4\) and \(t_{\text{max}}=25\). Three of the four players are aspirational CC agents, and one player is a nudging DRL agent. The sequential decisions of the players can be described as a Markov Decision Process (MDP), which is made up of a set of states \(S\), a set of actions \(A\), a transition probability function \(P(s, s')\), and a reward function \(R(s, s')\). The set of states \(S\) is defined as the contribution \(a_t\) each agent can make at time period \(t\).

At the onset of each round \(t\), every player receives one token. When taking an action, they can decide to contribute a fraction of this token to the collective pool, retaining the remainder. In a particular round \(t\), a player \(i\) can make a contribution \(a_{it}\) between 0 and 1. For simplicity, we assume that the \(a_{it}\) is continuous. After each round \(t\), the contributions in the pool are multiplied by a factor of \(k = 1.6\), following prior public goods game models \citep{ezaki_reinforcement_2016,horita_reinforcement_2017}. This amount is then evenly redistributed among all the players. Consequently, the reward \(r_{it}\), or payoff, an individual player \(i\) receives in round \(t\) is given by Equation 1:

\begin{equation}
r_{it} = \frac{k}{N}\sum_{j = 1}^{N}a_{jt} + (1 - a_{it})
\end{equation}
\label{eq:reward}

\subsection{The Reinforcement Learning Agents}\label{subsec2.2}

\subsubsection{The Aspirational Reinforcement Learning Agents}\label{subsubsec2.2.1}

We model our human CCs as aspirational RL agents following the Bush-Mosteller (BM) model of reinforcement learning \citep{ezaki_reinforcement_2016,mosteller_stochastic_1957}, a probabilistic rule-based learning model (see Section \ref{sec:related work}). BM agents do not seek to optimize the reward but seek to meet their aspirations using heuristics. The BM model is given in Equation \ref{eq:2}.

\begin{equation}
p_t =
    \begin{cases}
        p_{t - 1} + (1 - p_{t - 1})s_{t - 1} & \text{if } a_{t - 1} \geq X \text{ and } s_{t - 1} \geq 0 \\
        p_{t - 1} + p_{t - 1}s_{t - 1} & \text{if } a_{t - 1} \geq  X \text{ and } s_{t - 1} < 0 \\
        p_{t - 1} - (1 - p_{t - 1})s_{t - 1} & \text{if } a_{t - 1} < X \text{ and } s_{t - 1} \geq 0 \\
        p_{t - 1} - p_{t - 1}s_{t - 1} & \text{if } a_{t - 1}  < X \text{ and } s_{t - 1} < 0 \\
    \end{cases}
\label{eq:2}
\end{equation}

where \(p_t\) is the expected cooperative contribution that the player makes in round \(t\), and \(a_{t-1}\) is the action in \(t-1\), and \(s_{t-1}\) is the stimulus that drives learning (\(-1 < s_{t-1} < 1)\). The current action is suppressed if \(s_{t-1} < 0\), and reinforced if \(s_{t-1} > 0\), respectively. The stimulus is defined as: 
\begin{equation}
s_{t - 1} = \text{tanh}[\beta(r_{t - 1} - A)] 
\end{equation}

where \(r_{t - 1}\) is the reward (payoff) gained by a player in round \(t - 1\). \(A\) is the aspiration level (assumed to be fixed), and \(\beta\) is a parameter governing an agent's sensitivity to \(r_{t - 1} - A\). \(X\) is a threshold used to judge whether an agent is cooperative or not. Once \(p_t\) is computed, the actual contribution \(a_t\) is drawn from a Gaussian distribution with mean \(\mu = p_t\) and standard deviation \(\sigma = 0.2\). The first contribution is made randomly from a uniform distribution. In this paper we choose \(A = 1.0\), \(X = 0.4\), \(\beta = 0.4\), and \(\sigma = 0.2\). The choices of hyperparameters follow the findings in \cite{ezaki_reinforcement_2016}, where the CC models were "robustly observed if $\beta$ is larger than $\approx 0.2$, $A \leq 1$, and $0.1 \leq X \leq 0.4$". The actual parameter values themselves were chosen to be similar to hyperparameters used in \cite{ezaki_reinforcement_2016}.


\subsubsection{The Nudging Deep Reinforcement Learning (DRL) Agent}\label{subsubsec2.2.2}


Our social planner who seeks to nudge the CCs agents into cooperation is implemented using deep reinforcement learning (DRL).  Just like the aspiration-RL agent the DRL agent can take a set of actions \(A\) between 0 and 1 (for computational efficiency we discretized the actions for the DRL agents with increments of 0.01). The DRL agent seeks to optimize cooperation by learning a policy \(\pi(s, a)\) for its MDP process\citep{schulman_proximal_2017}. We model two different DRL agents, each seeking to optimize different reward functions. The first one seeks to optimize the total sum of the contributions $a_{it}$ of all $N-1$ CC agents (sum DRL). Its reward function is defined as (with the DRL agent denoted with index $j$): 

\begin{equation}
    R_{sum}(s_t, a) =
            \sum_{i = 1,i \neq j}^{N}a_{it}
\end{equation}

The second DRL agents seeks to optimize the proportion of the CC contributions $a_{it} > 0.5$ (prop DRL).

\begin{equation}
R_{prop}(s_t, a) =
       \frac{1}{N}\sum_{i = 1, i \neq j}^{N}(a_{it} > 0.5)
\end{equation}

\noindent{Our DRL agents learn the corresponding standard policy \(\pi(s, a)\) to maximize the cumulative reward over the period of the game with \(t_{\text{max}} = 25\): 
\begin{equation}
    \mathbb{E}_\pi[s, a] = \sum_{t=1}^{t_{\text{max}}} R_{sum}(s_t, a) \quad \text{or} \quad \mathbb{E}_\pi[s, a] = \sum_{t=1}^{t_{\text{max}}} R_{prop}(s_t, a)
\end{equation}
}


We used the PPO algorithm \citep{schulman_proximal_2017} because it follows an actor-critic structure. Prior work suggest that PPO is efficient for cooperative MARL \citep{hughes_inequity_2018,yu_surprising_2022}. 
The update equation for PPO is given by:

$$
L(\theta) = \hat{\mathbb{E}}_t \left[ \min \left( r_t(\theta) \hat{A}_t, \text{clip}(r_t(\theta), 1-\epsilon, 1+\epsilon) \hat{A}_t \right) \right],
$$

where \(r_t(\theta)\) is the probability ratio between the new and old policy, \(\hat{A}_t\) is the advantage estimate, and \(\epsilon\) is a hyperparameter controlling the clipping. 


\subsubsection{The Experiment Setup}\label{subsubsec2.2.3}

To evaluate the performance of our DRL agents as a social planner, we designed a set of experiments where we compare baseline games with CC agents only to games that contain a social planner that indirectly nudges the CC agents. Our social planner is a DRL agent (see Section \ref{subsubsec2.2.2})
\begin{table}[h]
\fontsize{9pt}{10pt}\selectfont
\centering
\caption{Experimental Groups}\label{experiment_groups_table}%
\vspace{2mm}
\begin{tabular}{@{}llll@{}}
\toprule
Group & Composition \\ \midrule
Baseline Games & 4 CC agents using aspiration-based RL\\ 
Nudging Games with Sum DRL & 3 CC agents, 1 sum DRL agent \\ 
Nuding Games with Prop DRL & 3 CC agents, 1 prop DRL agent \\ 
\bottomrule
\end{tabular}
\end{table}

We trained our DRL agents for a fixed 4,000,000 steps, consisting of 1,000 batches with 4,000 steps each, within a game environment that precisely mirrors their respective experimental group. During this training phase, the DRL agents aim to maximize their predefined reward function while playing with three CC agents. Following the training, each DRL agent participates in 10,000 evaluation games with three CC agents. In the baseline game, four CC agent engage in 10,000 evaluation games. The outcomes of these evaluation games are the primary focus of our study. The hyperparameter selections for the DRL agents are listed in Appendix A, Table \ref{tab:parameters}.

\section{Results}\label{sec2}

\subsection{Validation}

\begin{figure}[h]
\centering

\caption{The relationship between three CC agents' contributions in round $t - 1$ and the fourth CC agents' contribution in round $t$ for CC agents in the prior study \cite{ezaki_reinforcement_2016} (a) and in this paper (b).}

\includegraphics[scale=0.6]{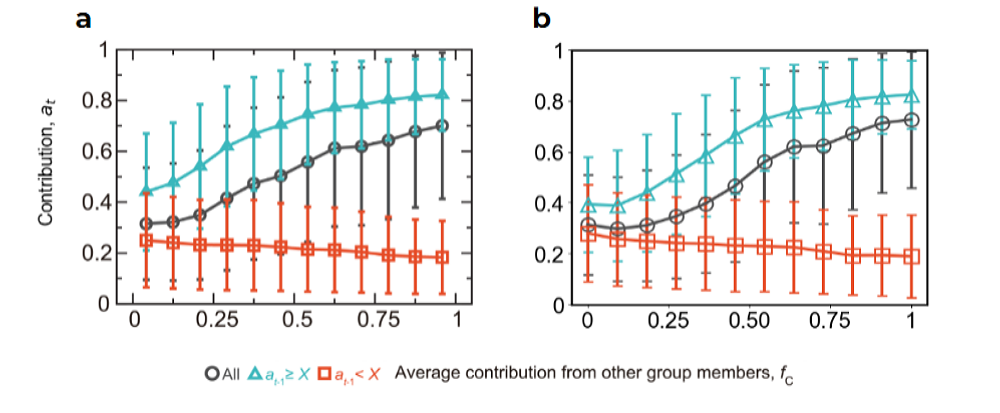}

\label{fig:validation}
\end{figure}

In Figure \ref{fig:validation}, we examine the empirical relationship between the contributions of three CC agents in round $t - 1$ and the contribution of the fourth CC agent in round $t$. The charts illustrate that CC agents typically respond to high average group contributions in the prior round with high contributions of their own, and to low average group contributions with low contributions. Additionally, the self-reinforcing nature of CC agents is evident: if an agent's prior contribution was below $X = 0.4$, they generally follow up any group contribution with a low contribution, and with a high contribution if their prior contribution was above $X$.). The plotted lines allow us to compare our findings to previous empirical validations \citep{ezaki_reinforcement_2016, horita_reinforcement_2017}, showing consistency with prior studies using empirical data.

\subsection{Deep Reinforcement Learning Agent Training}

\begin{figure}[h]
\centering
\caption{Mean episodic reward for both the sum (a) and prop (b) DRL agents throughout the training period of 4,000,000 games.}
\includegraphics[scale=0.25]{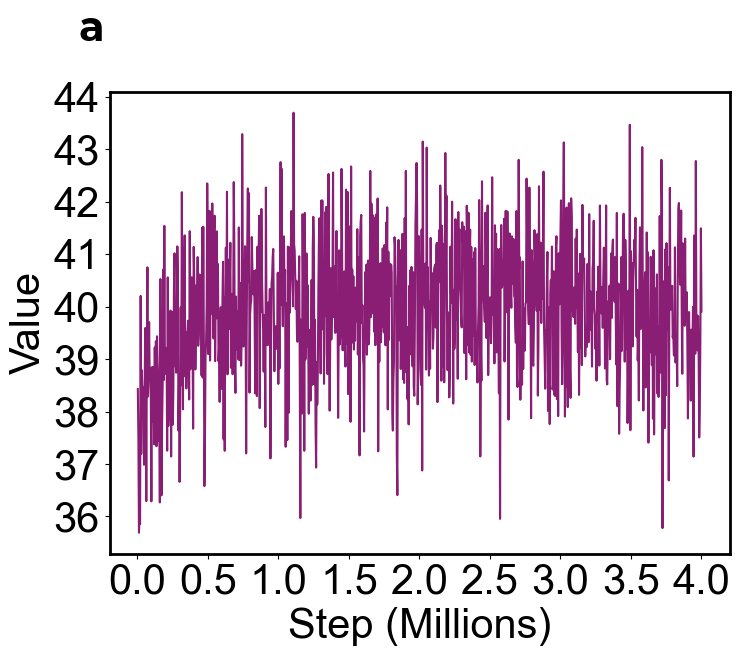} \hspace{0.5cm}
\includegraphics[scale=0.25]{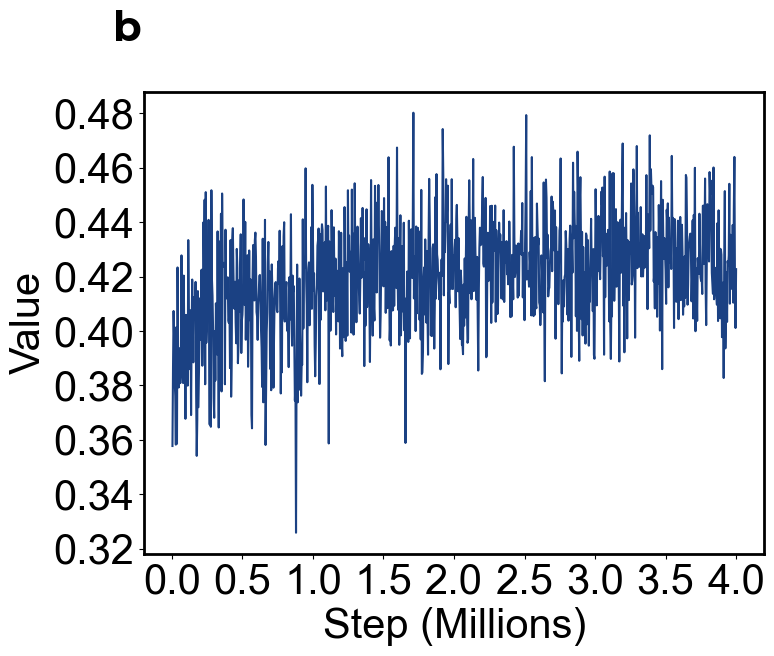}
\label{fig:episodic_reward}
\end{figure}

Figure \ref{fig:episodic_reward} present the training results for our deep RL agents. A step presents one game. The upward trends in Figures \ref{fig:episodic_reward} a and \ref{fig:episodic_reward} b indicate that each agent's performance and efficiency in learning optimal strategies improved over time. This consistent rise in rewards reflects the agents' learned ability to adapt and optimize their decision-making processes as training progressed.

\subsection{Game Results}

We find that both the sum DRL and the prop DRL agent were able to increase the sums of contributions and the proportion of contributions over 0.5 by notable amounts compared to the baseline. We performed a Mann-Whitney U-test to examine the statistical difference. The results are significant (see Table \ref{tab1}). 

In our games with 3 CC RL agents and one nudging RL agent, the sum DRL nudging agent increases the total sum of contributions by 8.22\% and the total proportion of cooperative contributions by 12.42\% compared to the baseline. In our games with 3 CC RL agents and one nudging RL agent, the prop DRL nudging agent increases the total sum of contributions by 8.85\% and the total proportion of cooperative contributions by 14.87\%, compared to the baseline. While it appears that the prop DRL agent is more successful in nudging CC behavior, the difference is not statistically significant (The Mann-Whitney U-test when comparing the two distributions of game outcomes was not significant; $p = 0.139$ (Prop Contribution >0.5) and $p = 0.766$ (Sum Contribution Mean)).

\begin{table}[h]
\fontsize{9pt}{10pt}\selectfont
\centering
\caption{Game Outcomes}\label{tab1}%
\vspace{2mm}
\begin{tabular}{@{}lllll@{}}
\toprule
Measure & Baseline & Sum DRL & Prop DRL \\ \midrule
Sum Contribution (mean) & 36.995 & 40.035 & 40.268\\ 
Mann-Whitney U-test &  & 1.042 $\cdot$ 10$^{-44}$ &  1.318 $\cdot$ 10$^{-58}$\\
Percentage Change & & 8.22\% & 8.85\% \\ \midrule
Prop. Contribution > 0.5 (mean) & 0.491 & 0.552 & 0.56\\ 
Mann-Whitney U-test & & 1.318 $\cdot$ 10$^{-58}$ &  1.439 $\cdot$ 10$^{-64}$ \\  
Percentage Change & & 12.42\% & 14.87\% \\ 
\bottomrule
\end{tabular}
\end{table}


Overall, the results confirm that both DRL agents are successfully nudging CC. They are foster cooperation with a significant increase over the total period of the game. 


\begin{figure}[h]
\centering
\includegraphics[scale=0.24]{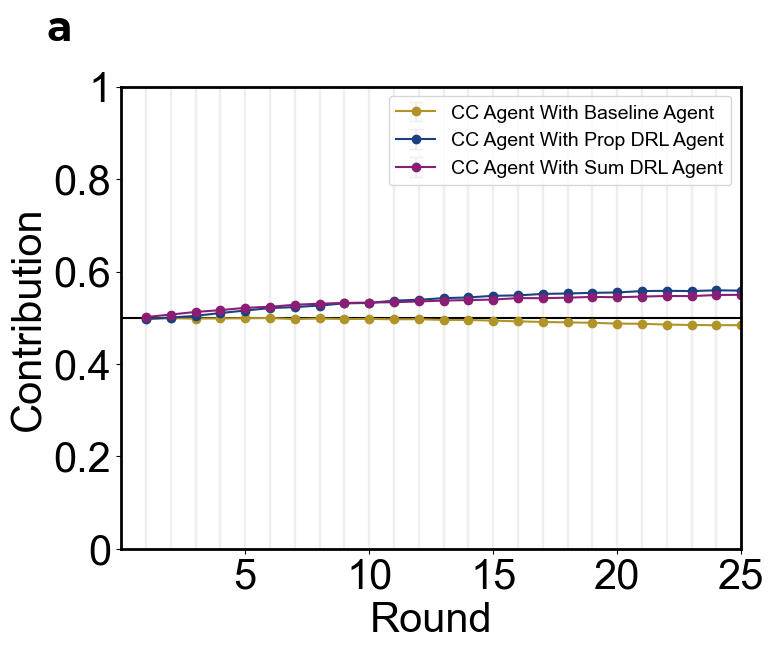}
\includegraphics[scale=0.24]{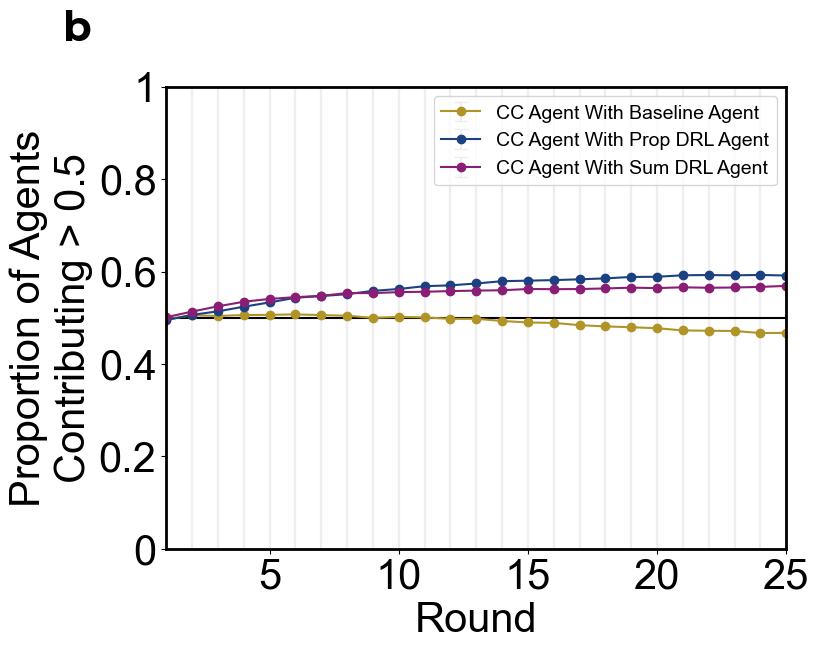}
\includegraphics[scale=0.24]{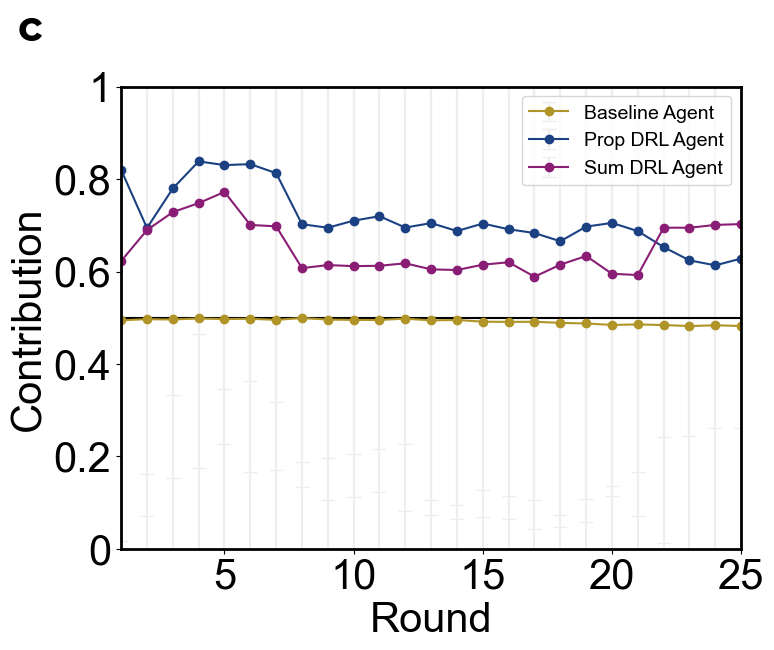}
\caption{The mean contributions of the three other CC agents (a), the proportion of their contributions deemed cooperative when interacting with baseline agents and the DRL agents (b), and the average contributions of both baseline and DRL agents across rounds (c).}
\label{fig:contributions}
\end{figure}

Figures \ref{fig:contributions}a, \ref{fig:contributions}b, and \ref{fig:contributions}c are based on taking the mean per-round contributions across 10,000 evaluation games ran after the DRL agents were finished training. Notably, the DRL agents diverge from the baseline agents' trend early on due to their substantial contributions in the initial seven rounds, which surpass those of the baseline agents. Once a social norm is established, both DRL agents stabilize their contributions at consistently higher levels than the control agents. This early establishment of norms by the DRL agents encourages the CC agents to assume the responsibility of maintaining high contributions, thereby allowing the DRL agents to reduce their contributions while still benefiting from the established norm.

\begin{figure}[h]
\centering
\includegraphics[scale=0.18]{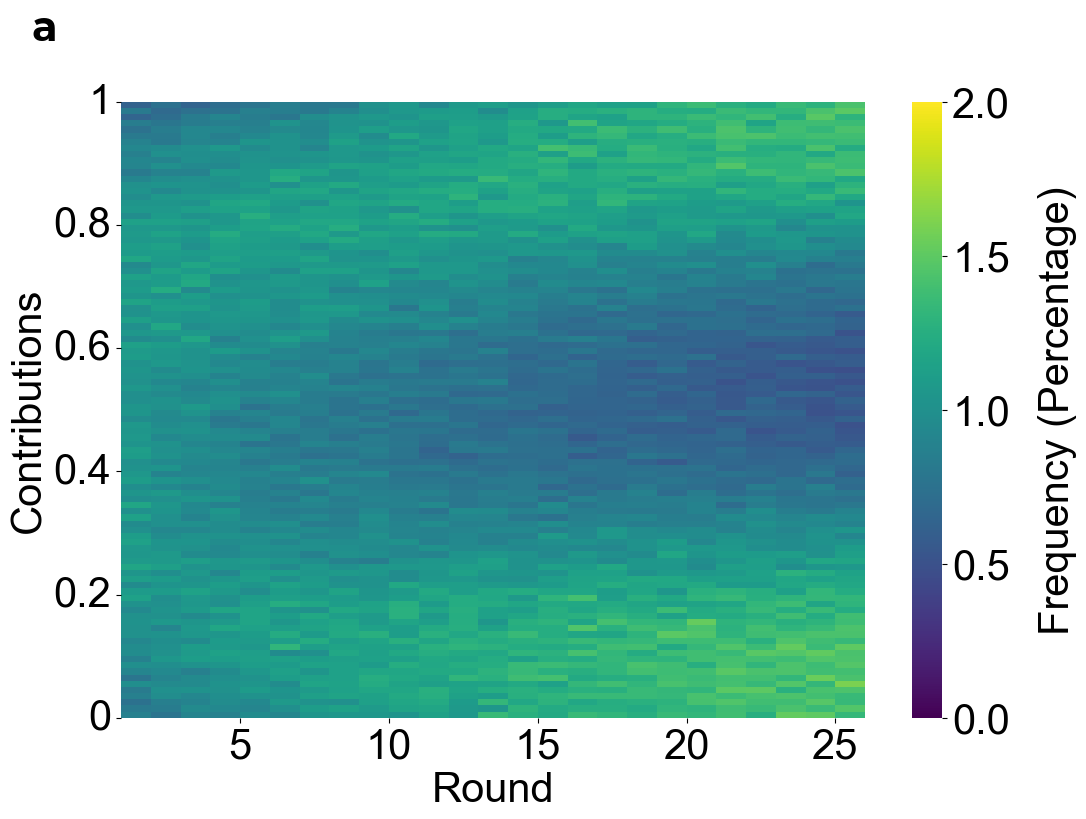}
\includegraphics[scale=0.18]{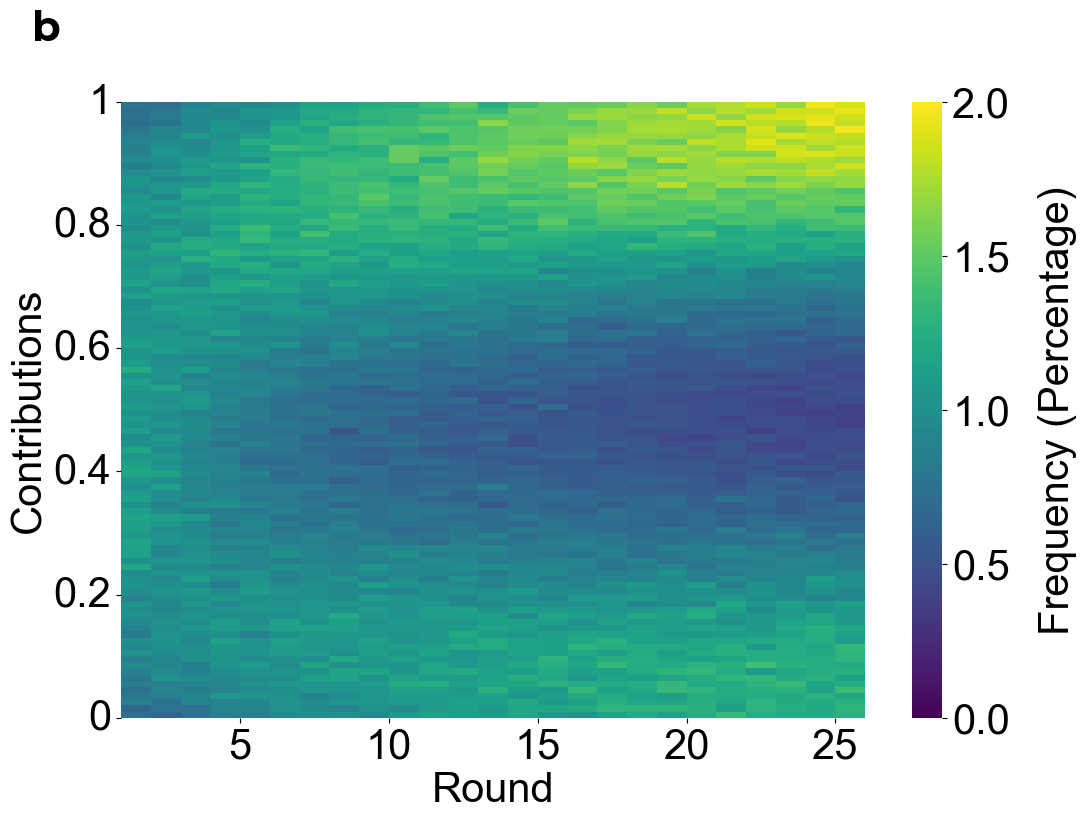}
\includegraphics[scale=0.18]{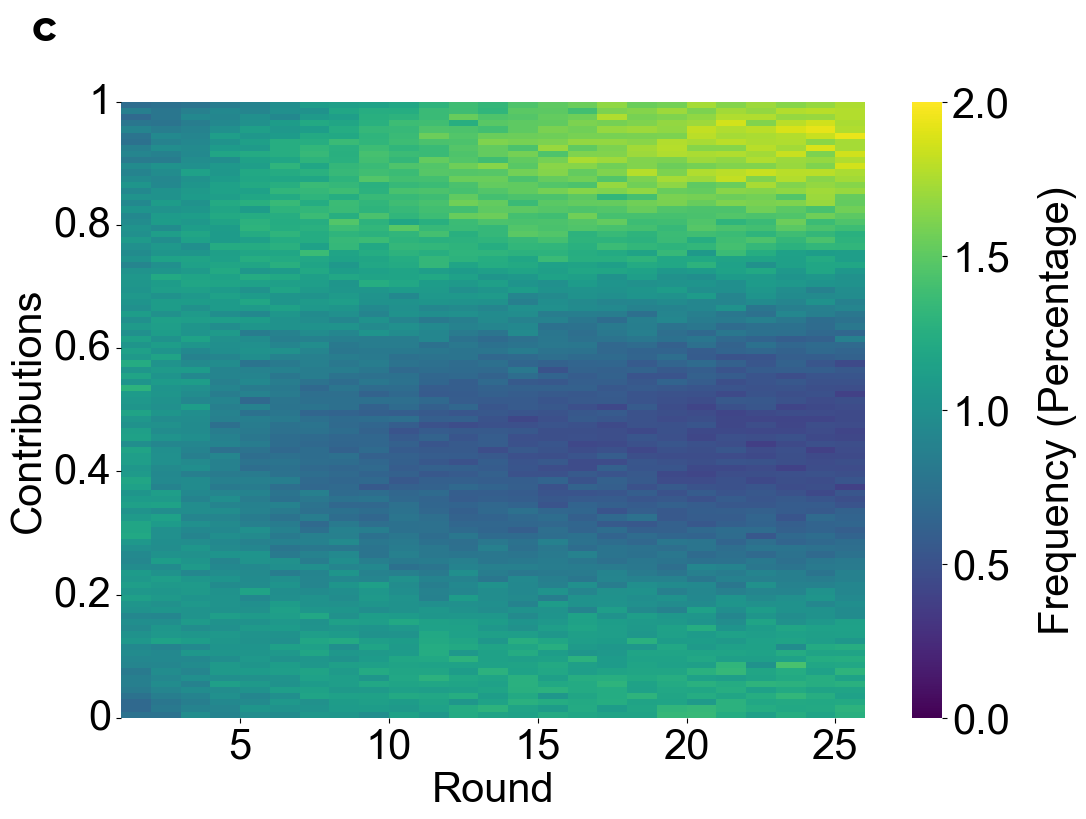}
\caption{The distributions of CC agents when playing with the baseline (a), the sum DRL agent (b), and the prop DRL agent (c) at each round represented with heat maps.}
\label{fig:contribution_distributions}
\end{figure}

Figures \ref{fig:contribution_distributions}a and \ref{fig:contribution_distributions}b show a tendency for CC agents to follow one of two paths: either the low-contribution "doom loops," where the CC agents start contributing low and end up contributing even lower as the game progresses, or positive-feedback loops, where the contributions trend upward as the game progresses. The heat maps for the baseline agent group show a general trend towards the "doom loop" scenario, suggesting that the baseline (CC) agent is not very effective at preventing these situations from occurring. In contrast, the heat maps for both the prop DRL and sum DRL agents show a noticeable shift towards higher contributions at the later rounds, suggesting that the agents' strategy of contributing early does significantly increase the odds of creating these positive-feedback loops, as evidenced by the increased brightness of the upper path across the later rounds. Of interesting note is that the CC agents' contributions begin trending upward notably around round 10, well after the DRL agents drop their contributions. This is likely the point at which the positive reinforcement loop is established and the CC agents themselves have taken over upholding the cooperative social norm.

\begin{figure}[h]
\centering
\includegraphics[scale=0.24]{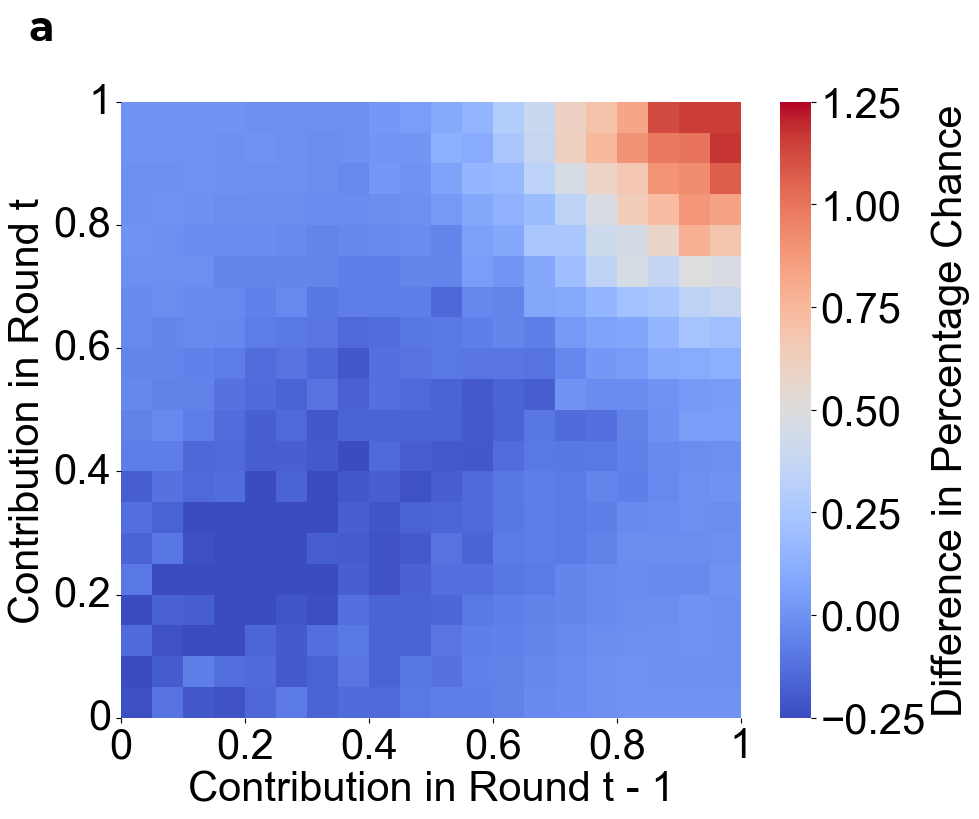} \hspace{0.5cm}
\includegraphics[scale=0.24]{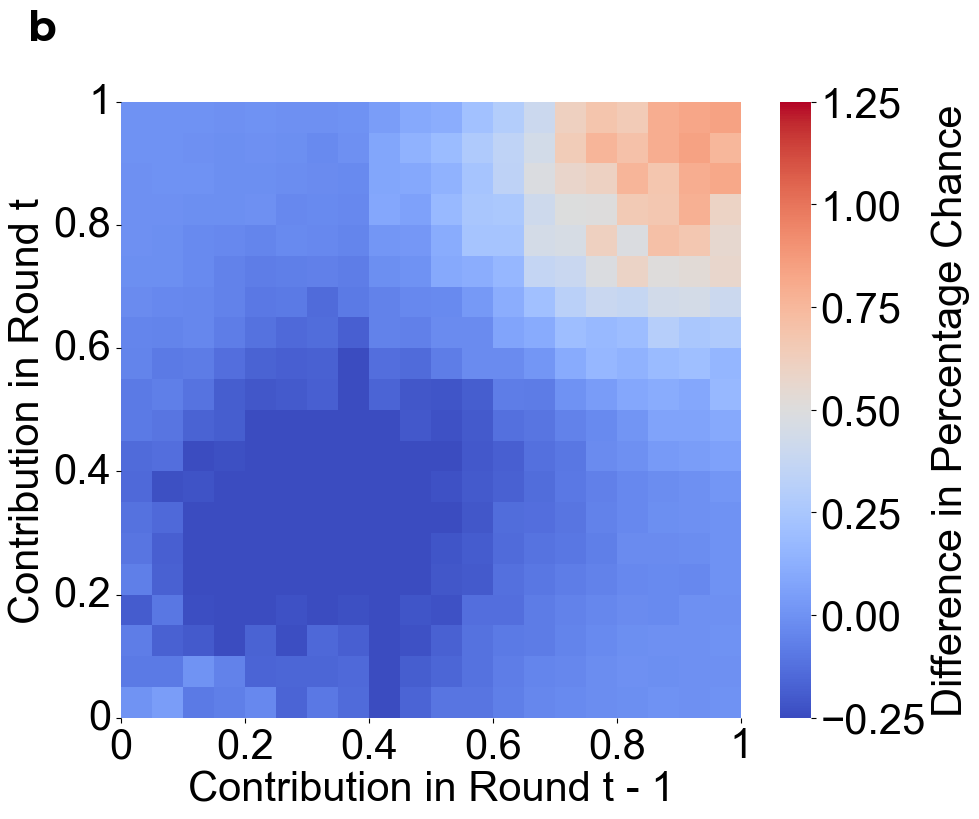}
\caption{The difference in frequency of contributions by CC agents when playing with the sum DRL agent (a) and prop DRL agent (b) compared to baseline agents. These heatmaps show the change in probability of following up contributions with higher or lower amounts, indicating the positive influence of both DRL agents on cooperative behavior.}
\label{fig:ai_difference_heatmaps}
\end{figure}

Figures \ref{fig:ai_difference_heatmaps}a and \ref{fig:ai_difference_heatmaps}b demonstrate the increased confidence that CC agents gain when contributing higher amounts due to these agents and also the lower probabilities of following low contributions with other low contributions. Taken together, these charts show the positive influence both DRL agents exert upon the CC agents they play with. The detailed analysis of these heat maps shows that both DRL agents have a pronounced effect on encouraging high contributions when compared to the baseline agent. The DRL agents' influence is particularly notable in the increase in probability of CC agents following up a high contribution with another high contribution along with the decrease in probability of following up a low contributions with another low contribution. This generally shows that the DRL agents have done a good job with getting CC agents to avoid doubling down on negative trends and reinforcing positive trends.

When piecing all of these charts together, we can clearly see that the strategy that both DRL agents focused on was to set a social norm that the CC agents felt empowered to contribute cooperatively in, and continue to nurture that over the course of the game. By this measure, both DRL agents were similarly successful, despite some variation in strategies. The sum DRL agent seems to do a better job at incentivizing CC agents to follow up high contributions with more high contributions (Fig. \ref{fig:ai_difference_heatmaps}) However, the prop DRL agent creates a slightly more robust distribution of contributions by CC agents in the later round (Fig. \ref{fig:contribution_distributions}), indicating that CC agents playing with the prop DRL agent tend to create the positive feedback loop slightly more often. Nevertheless, the differences in the two DRL agents pale in comparison to the differences between each agent and the baseline, with both DRL agents demonstrating a significant positive influence on the proceedings of the game after round 10 that the baseline agent was unable to replicate (Fig. \ref{fig:contribution_distributions}). The key reason the baseline agent could not create a trend is that it is also a CC agent, and as a result reacts to trends in a way that DRL agents with personal incentives will not. This aligns with prior research regarding social norms and conditional cooperators \citep{te_velde_conformity_2022, cialdini_descriptive_2007} suggesting that non-CC actors are able to establish norms which CC agents will follow.

\section{Conclusion}\label{sec2}

Our findings reveal that both DRL agents outperformed the baseline model consisting solely of CC agents. The sum DRL agent increased the total contributions by 8.22\% and the proportion of cooperative contributions by 12.42\%, while the prop DRL agent increased these metrics by 8.85\% and 14.87\%, respectively.

The results indicate that the presence of a nudging agent significantly influences the behavior of CC agents, encouraging them to adopt higher levels of cooperation. This extends prior research on social norms with a machine learning perspective \citep{keser_conditional_2000, cialdini_descriptive_2007}; deep RL agents acting as a "social planner" with an invisible hand can encourage cooperative behavior through social incentives. 

The strategies used by both DRL agents, despite their different reward functions, suggest that the early establishment of a cooperative norm is crucial. By contributing significantly in the initial rounds, the DRL agents set a precedent that the CC agents follow, thereby creating a positive feedback loop of cooperation. This finding underscores the importance of early intervention in promoting collective action, a concept well-documented in the literature on social dilemmas \citep{keser_conditional_2000}.

In the machine learning space, our research builds on existing work in multi-agent reinforcement learning (MARL) and cooperative behavior \citep{jaques_social_2019, gronauer_multi-agent_2022}. By introducing a nudging DRL agent, we add to the understanding of how decentralized policies can be optimized to promote collective action. From a social science perspective, our study contributes to the literature on public goods games and conditional cooperation \citep{ezaki_reinforcement_2016, horita_reinforcement_2017, ledyard_public_1995}. By integrating aspiration-based reinforcement learning with social norm nudges, we provide a novel approach to understanding and influencing cooperative behavior. Our findings support the notion that strategic interventions can foster cooperation, offering practical insights for policymakers and researchers interested in promoting collective action in various societal contexts \citep{te_velde_conformity_2022, cialdini_descriptive_2007}.

While our model shows promising results, it has certain limitations. Firstly, the homogeneity of the CC agents in our model does not account for the diversity of behaviors observed in real-world scenarios, where free riders and unconditional cooperators also play a role. This research also did not explore different variants of the public goods game, such as single-shot PGG. Future research should explore these dimensions to fully understand the applicability and limitations of AI-guided cooperation in diverse settings.

In conclusion, this study accentuates the profound potential of DRL models in optimizing cooperation within the public goods game. It paves the way for future explorations into AI-guided social norms and their role in guiding conditionally cooperative actors toward collective action. Our findings serve as a pivotal reference, illustrating how AI can shape social norms and foster cooperation, thereby contributing significantly to both machine learning literature and social science discourse.

\bibliographystyle{plainnat}  
\bibliography{references}  

\begin{thebibliography}{34}
\providecommand{\natexlab}[1]{#1}
\providecommand{\url}[1]{\texttt{#1}}
\expandafter\ifx\csname urlstyle\endcsname\relax
  \providecommand{\doi}[1]{doi: #1}\else
  \providecommand{\doi}{doi: \begingroup \urlstyle{rm}\Url}\fi

\bibitem[Ackermann and Murphy(2019)]{ackermann_explaining_2019}
Kurt Ackermann and Ryan Murphy.
\newblock Explaining {Cooperative} {Behavior} in {Public} {Goods} {Games}: {How} {Preferences} and {Beliefs} {Affect} {Contribution} {Levels}.
\newblock \emph{Games}, 10\penalty0 (1):\penalty0 15, March 2019.
\newblock ISSN 2073-4336.
\newblock \doi{10.3390/g10010015}.
\newblock URL \url{https://www.mdpi.com/2073-4336/10/1/15}.

\bibitem[Andı and Akesson(2021)]{andi_nudging_2021}
Simge Andı and Jesper Akesson.
\newblock Nudging {Away} {False} {News}: {Evidence} from a {Social} {Norms} {Experiment}.
\newblock \emph{Digital Journalism}, 9\penalty0 (1):\penalty0 106--125, January 2021.
\newblock ISSN 2167-0811.
\newblock \doi{10.1080/21670811.2020.1847674}.
\newblock URL \url{https://doi.org/10.1080/21670811.2020.1847674}.
\newblock Publisher: Routledge \_eprint: https://doi.org/10.1080/21670811.2020.1847674.

\bibitem[Bendor et~al.(2001)Bendor, Mookherjee, and Ray]{bendor_aspiration-based_2001}
Jonathan Bendor, Dilip Mookherjee, and Debraj Ray.
\newblock Aspiration-based reinforcement learning in repeated interaction games: an overview.
\newblock \emph{International Game Theory Review}, 03\penalty0 (02n03):\penalty0 159--174, June 2001.
\newblock ISSN 0219-1989.
\newblock \doi{10.1142/S0219198901000348}.
\newblock URL \url{https://www.worldscientific.com/doi/abs/10.1142/S0219198901000348}.
\newblock Publisher: World Scientific Publishing Co.

\bibitem[Canese et~al.(2021)Canese, Cardarilli, Di~Nunzio, Fazzolari, Giardino, Re, and Spanò]{canese_multi-agent_2021}
Lorenzo Canese, Gian~Carlo Cardarilli, Luca Di~Nunzio, Rocco Fazzolari, Daniele Giardino, Marco Re, and Sergio Spanò.
\newblock Multi-{Agent} {Reinforcement} {Learning}: {A} {Review} of {Challenges} and {Applications}.
\newblock \emph{Applied Sciences}, 11\penalty0 (11):\penalty0 4948, January 2021.
\newblock ISSN 2076-3417.
\newblock \doi{10.3390/app11114948}.
\newblock URL \url{https://www.mdpi.com/2076-3417/11/11/4948}.
\newblock Number: 11 Publisher: Multidisciplinary Digital Publishing Institute.

\bibitem[Cialdini(2007)]{cialdini_descriptive_2007}
Robert~B. Cialdini.
\newblock Descriptive {Social} {Norms} as {Underappreciated} {Sources} of {Social} {Control}.
\newblock \emph{Psychometrika}, 72\penalty0 (2):\penalty0 263--268, June 2007.
\newblock ISSN 1860-0980.
\newblock \doi{10.1007/s11336-006-1560-6}.
\newblock URL \url{https://doi.org/10.1007/s11336-006-1560-6}.

\bibitem[Cross(1973)]{cross_stochastic_1973}
John~G. Cross.
\newblock A {Stochastic} {Learning} {Model} of {Economic} {Behavior}*.
\newblock \emph{The Quarterly Journal of Economics}, 87\penalty0 (2):\penalty0 239--266, May 1973.
\newblock ISSN 0033-5533.
\newblock \doi{10.2307/1882186}.
\newblock URL \url{https://doi.org/10.2307/1882186}.

\bibitem[Ezaki et~al.(2016)Ezaki, Horita, Takezawa, and Masuda]{ezaki_reinforcement_2016}
Takahiro Ezaki, Yutaka Horita, Masanori Takezawa, and Naoki Masuda.
\newblock Reinforcement {Learning} {Explains} {Conditional} {Cooperation} and {Its} {Moody} {Cousin}.
\newblock \emph{PLOS Computational Biology}, 12\penalty0 (7):\penalty0 e1005034, July 2016.
\newblock ISSN 1553-7358.
\newblock \doi{10.1371/journal.pcbi.1005034}.
\newblock URL \url{https://journals.plos.org/ploscompbiol/article?id=10.1371/journal.pcbi.1005034}.
\newblock Publisher: Public Library of Science.

\bibitem[Fischbacher et~al.(2001)Fischbacher, Gächter, and Fehr]{fischbacher_are_2001}
Urs Fischbacher, Simon Gächter, and Ernst Fehr.
\newblock Are people conditionally cooperative? {Evidence} from a public goods experiment.
\newblock \emph{Economics Letters}, 71\penalty0 (3):\penalty0 397--404, June 2001.
\newblock ISSN 0165-1765.
\newblock \doi{10.1016/S0165-1765(01)00394-9}.
\newblock URL \url{https://www.sciencedirect.com/science/article/pii/S0165176501003949}.

\bibitem[Gronauer and Diepold(2022)]{gronauer_multi-agent_2022}
Sven Gronauer and Klaus Diepold.
\newblock Multi-agent deep reinforcement learning: a survey.
\newblock \emph{Artificial Intelligence Review}, 55\penalty0 (2):\penalty0 895--943, February 2022.
\newblock ISSN 1573-7462.
\newblock \doi{10.1007/s10462-021-09996-w}.
\newblock URL \url{https://doi.org/10.1007/s10462-021-09996-w}.

\bibitem[Hardin(1968)]{hardin_tragedy_1968}
Garrett Hardin.
\newblock The {Tragedy} of the {Commons}.
\newblock \emph{Science}, 162\penalty0 (3859):\penalty0 1243--1248, December 1968.
\newblock \doi{10.1126/science.162.3859.1243}.
\newblock URL \url{https://www.science.org/doi/10.1126/science.162.3859.1243}.
\newblock Publisher: American Association for the Advancement of Science.

\bibitem[Horita et~al.(2017)Horita, Takezawa, Inukai, Kita, and Masuda]{horita_reinforcement_2017}
Yutaka Horita, Masanori Takezawa, Keigo Inukai, Toshimasa Kita, and Naoki Masuda.
\newblock Reinforcement learning accounts for moody conditional cooperation behavior: experimental results.
\newblock \emph{Scientific Reports}, 7\penalty0 (1):\penalty0 39275, January 2017.
\newblock ISSN 2045-2322.
\newblock \doi{10.1038/srep39275}.
\newblock URL \url{https://www.nature.com/articles/srep39275}.
\newblock Bandiera\_abtest: a Cc\_license\_type: cc\_by Cg\_type: Nature Research Journals Number: 1 Primary\_atype: Research Publisher: Nature Publishing Group Subject\_term: Human behaviour;Social evolution Subject\_term\_id: human-behaviour;social-evolution.

\bibitem[Hughes et~al.(2018)Hughes, Leibo, Phillips, Tuyls, Dueñez-Guzman, García~Castañeda, Dunning, Zhu, McKee, Koster, Roff, and Graepel]{hughes_inequity_2018}
Edward Hughes, Joel~Z Leibo, Matthew Phillips, Karl Tuyls, Edgar Dueñez-Guzman, Antonio García~Castañeda, Iain Dunning, Tina Zhu, Kevin McKee, Raphael Koster, Heather Roff, and Thore Graepel.
\newblock Inequity aversion improves cooperation in intertemporal social dilemmas.
\newblock In \emph{Advances in {Neural} {Information} {Processing} {Systems}}, volume~31. Curran Associates, Inc., 2018.
\newblock URL \url{https://papers.nips.cc/paper/2018/hash/7fea637fd6d02b8f0adf6f7dc36aed93-Abstract.html}.

\bibitem[Jaques et~al.(2019)Jaques, Lazaridou, Hughes, Gulcehre, Ortega, Strouse, Leibo, and Freitas]{jaques_social_2019}
Natasha Jaques, Angeliki Lazaridou, Edward Hughes, Caglar Gulcehre, Pedro Ortega, Dj~Strouse, Joel~Z. Leibo, and Nando~De Freitas.
\newblock Social {Influence} as {Intrinsic} {Motivation} for {Multi}-{Agent} {Deep} {Reinforcement} {Learning}.
\newblock In \emph{Proceedings of the 36th {International} {Conference} on {Machine} {Learning}}, pages 3040--3049. PMLR, May 2019.
\newblock URL \url{https://proceedings.mlr.press/v97/jaques19a.html}.
\newblock ISSN: 2640-3498.

\bibitem[Jentoft et~al.(2018)Jentoft, Bavinck, Alonso-Población, Child, Diegues, Kalikoski, Kurien, McConney, Onyango, Siar, and Rivera]{jentoft_working_2018}
Svein Jentoft, Maarten Bavinck, Enrique Alonso-Población, Anna Child, Antonio Diegues, Daniela Kalikoski, John Kurien, Patrick McConney, Paul Onyango, Susana Siar, and Vivienne~Solis Rivera.
\newblock Working together in small-scale fisheries: harnessing collective action for poverty eradication.
\newblock \emph{Maritime Studies}, 17\penalty0 (1):\penalty0 1--12, April 2018.
\newblock ISSN 1872-7859, 2212-9790.
\newblock \doi{10.1007/s40152-018-0094-8}.
\newblock URL \url{http://link.springer.com/10.1007/s40152-018-0094-8}.

\bibitem[Keser and Van~Winden(2000)]{keser_conditional_2000}
Claudia Keser and Frans Van~Winden.
\newblock Conditional {Cooperation} and {Voluntary} {Contributions} to {Public} {Goods}.
\newblock \emph{The Scandinavian Journal of Economics}, 102\penalty0 (1):\penalty0 23--39, March 2000.
\newblock ISSN 0347-0520, 1467-9442.
\newblock \doi{10.1111/1467-9442.00182}.
\newblock URL \url{https://onlinelibrary.wiley.com/doi/10.1111/1467-9442.00182}.

\bibitem[Lang et~al.(2018)Lang, DeAngelo, and Bongard]{lang_explaining_2018}
Hannes Lang, Gregory DeAngelo, and Michelle Bongard.
\newblock Explaining {Public} {Goods} {Game} {Contributions} with {Rational} {Ability}.
\newblock \emph{Games}, 9\penalty0 (2):\penalty0 36, June 2018.
\newblock ISSN 2073-4336.
\newblock \doi{10.3390/g9020036}.
\newblock URL \url{http://www.mdpi.com/2073-4336/9/2/36}.

\bibitem[Ledyard(1995)]{ledyard_public_1995}
John~O. Ledyard.
\newblock \emph{Public {Goods}: {A} {Survey} of {Experimental} {Research}}.
\newblock Princeton University Press, Princeton, N.J, 1995.
\newblock ISBN 9780691042909.

\bibitem[March(1988)]{march_variable_1988}
James~G. March.
\newblock Variable risk preferences and adaptive aspirations.
\newblock \emph{Journal of Economic Behavior \& Organization}, 9\penalty0 (1):\penalty0 5--24, January 1988.
\newblock ISSN 0167-2681.
\newblock \doi{10.1016/0167-2681(88)90004-2}.
\newblock URL \url{https://www.sciencedirect.com/science/article/pii/0167268188900042}.

\bibitem[McKee et~al.(2023)McKee, Tacchetti, Bakker, Balaguer, Campbell-Gillingham, Everett, and Botvinick]{mckee_scaffolding_2023}
Kevin~R. McKee, Andrea Tacchetti, Michiel~A. Bakker, Jan Balaguer, Lucy Campbell-Gillingham, Richard Everett, and Matthew Botvinick.
\newblock Scaffolding cooperation in human groups with deep reinforcement learning.
\newblock \emph{Nature Human Behaviour}, 7\penalty0 (10):\penalty0 1787--1796, October 2023.
\newblock ISSN 2397-3374.
\newblock \doi{10.1038/s41562-023-01686-7}.
\newblock URL \url{https://www.nature.com/articles/s41562-023-01686-7}.
\newblock Publisher: Nature Publishing Group.

\bibitem[Mosteller(1957)]{mosteller_stochastic_1957}
Frederick Mosteller.
\newblock Stochastic {Models} for the {Learning} {Process}.
\newblock \emph{Proceedings of the American Philosophical Society}, 1957.
\newblock URL \url{https://www.jstor.org/stable/985304}.

\bibitem[Munoz~de Cote et~al.(2006)Munoz~de Cote, Lazaric, and Restelli]{munoz_de_cote_learning_2006}
Enrique Munoz~de Cote, Alessandro Lazaric, and Marcello Restelli.
\newblock Learning to cooperate in multi-agent social dilemmas.
\newblock \emph{AAMAS}, pages 783--785, May 2006.
\newblock \doi{https://doi.org/10.1145/1160633.1160770}.
\newblock URL \url{https://dl.acm.org/doi/10.1145/1160633.1160770}.

\bibitem[Ostrom(2000)]{ostrom_elinor_conversation_2000}
DAVID Ostrom, Elinor.
\newblock Conversation and {Cooperation} in {Social} {Dilemmas}: {A} {Meta}-{Analysis} of {Experiments} from 1958 to 1992.
\newblock \emph{Rationality and Society}, 7\penalty0 (1):\penalty0 58--92, 2000.
\newblock ISSN 1043-4631.
\newblock \doi{10.1177/1043463195007001004}.
\newblock URL \url{https://doi.org/10.1177/1043463195007001004}.

\bibitem[Schulman et~al.(2017)Schulman, Wolski, Dhariwal, Radford, and Klimov]{schulman_proximal_2017}
John Schulman, Filip Wolski, Prafulla Dhariwal, Alec Radford, and Oleg Klimov.
\newblock Proximal {Policy} {Optimization} {Algorithms}, August 2017.
\newblock URL \url{http://arxiv.org/abs/1707.06347}.
\newblock arXiv:1707.06347 [cs].

\bibitem[Simon(1955)]{simon_behavioral_1955}
Herbert~A. Simon.
\newblock A {Behavioral} {Model} of {Rational} {Choice}.
\newblock \emph{The Quarterly Journal of Economics}, 69\penalty0 (1):\penalty0 99--118, February 1955.
\newblock ISSN 0033-5533.
\newblock \doi{10.2307/1884852}.
\newblock URL \url{https://academic.oup.com/qje/article/69/1/99/1919737}.

\bibitem[Simon et~al.(1987)Simon, Dantzig, Hogarth, Plott, Raiffa, Schelling, Shepsle, Thaler, Tversky, and Winter]{simon_decision_1987}
Herbert~A. Simon, George~B. Dantzig, Robin Hogarth, Charles~R. Plott, Howard Raiffa, Thomas~C. Schelling, Kenneth~A. Shepsle, Richard Thaler, Amos Tversky, and Sidney Winter.
\newblock Decision {Making} and {Problem} {Solving}.
\newblock \emph{Interfaces}, 17:\penalty0 11, October 1987.
\newblock ISSN 00922102.

\bibitem[Song et~al.(2022)Song, Guo, Jia, Perc, Li, and Wang]{song_reinforcement_2022}
Zhao Song, Hao Guo, Danyang Jia, Matjaž Perc, Xuelong Li, and Zhen Wang.
\newblock Reinforcement learning facilitates an optimal interaction intensity for cooperation.
\newblock \emph{Neurocomputing}, 513:\penalty0 104--113, November 2022.
\newblock ISSN 0925-2312.
\newblock \doi{10.1016/j.neucom.2022.09.109}.
\newblock URL \url{https://www.sciencedirect.com/science/article/pii/S0925231222012000}.

\bibitem[Stahl and Haruvy(2002)]{stahl_aspiration-based_2002}
Dale~O. Stahl and Ernan Haruvy.
\newblock Aspiration-{Based} and {Reciprocity}-{Based} {Rules} in {Learning} {Dynamics} for {Symmetric} {Normal}-{Form} {Games}.
\newblock \emph{Journal of Mathematical Psychology}, 46\penalty0 (5):\penalty0 531--553, October 2002.
\newblock ISSN 0022-2496.
\newblock \doi{10.1006/jmps.2001.1409}.
\newblock URL \url{https://www.sciencedirect.com/science/article/pii/S0022249601914099}.

\bibitem[Sutton and Barto(2020)]{sutton_reinforcement_2020}
Richard~S. Sutton and Andrew~G Barto.
\newblock \emph{Reinforcement {Learning}: {An} {Introduction}}.
\newblock MIT Press, 2 edition, 2020.
\newblock ISBN 978-0-262-19398-6.
\newblock URL \url{https://www.andrew.cmu.edu/course/10-703/textbook/BartoSutton.pdf}.

\bibitem[Tampuu et~al.(2017)Tampuu, Matiisen, Kodelja, Kuzovkin, Korjus, Aru, Aru, and Vicente]{tampuu_multiagent_2017}
Ardi Tampuu, Tambet Matiisen, Dorian Kodelja, Ilya Kuzovkin, Kristjan Korjus, Juhan Aru, Jaan Aru, and Raul Vicente.
\newblock Multiagent cooperation and competition with deep reinforcement learning.
\newblock \emph{PLOS ONE}, 12\penalty0 (4):\penalty0 e0172395, April 2017.
\newblock ISSN 1932-6203.
\newblock \doi{10.1371/journal.pone.0172395}.
\newblock URL \url{https://journals.plos.org/plosone/article?id=10.1371/journal.pone.0172395}.
\newblock Publisher: Public Library of Science.

\bibitem[te~Velde and Louis(2022)]{te_velde_conformity_2022}
Vera~L. te~Velde and Winnifred Louis.
\newblock Conformity to descriptive norms.
\newblock \emph{Journal of Economic Behavior \& Organization}, 200:\penalty0 204--222, August 2022.
\newblock ISSN 0167-2681.
\newblock \doi{10.1016/j.jebo.2022.05.017}.
\newblock URL \url{https://www.sciencedirect.com/science/article/pii/S0167268122001780}.

\bibitem[Tomassini and Antonioni(2021)]{tomassini_computational_2021}
Marco Tomassini and Alberto Antonioni.
\newblock Computational behavioral models in public goods games with migration between groups.
\newblock \emph{Journal of Physics: Complexity}, 2\penalty0 (4):\penalty0 045013, November 2021.
\newblock ISSN 2632-072X.
\newblock \doi{10.1088/2632-072X/ac371b}.
\newblock URL \url{https://dx.doi.org/10.1088/2632-072X/ac371b}.
\newblock Publisher: IOP Publishing.

\bibitem[Yang et~al.(2020)Yang, Li, Farajtabar, Sunehag, Hughes, and Zha]{yang_learning_2020}
Jiachen Yang, Ang Li, Mehrdad Farajtabar, Peter Sunehag, Edward Hughes, and Hongyuan Zha.
\newblock Learning to incentivize other learning agents.
\newblock In \emph{Proceedings of the 34th {International} {Conference} on {Neural} {Information} {Processing} {Systems}}, {NIPS} '20, pages 15208--15219, Red Hook, NY, USA, December 2020. Curran Associates Inc.
\newblock ISBN 978-1-71382-954-6.

\bibitem[Yu et~al.(2022)Yu, Velu, Vinitsky, Gao, Wang, Bayen, and Wu]{yu_surprising_2022}
Chao Yu, Akash Velu, Eugene Vinitsky, Jiaxuan Gao, Yu~Wang, Alexandre Bayen, and Yi~Wu.
\newblock The {Surprising} {Effectiveness} of {PPO} in {Cooperative} {Multi}-{Agent} {Games}.
\newblock \emph{Advances in Neural Information Processing Systems}, 35:\penalty0 24611--24624, December 2022.
\newblock URL \url{https://proceedings.neurips.cc/paper_files/paper/2022/hash/9c1535a02f0ce079433344e14d910597-Abstract-Datasets_and_Benchmarks.html}.

\bibitem[Zhang et~al.(2021)Zhang, Yang, and Başar]{zhang_multi-agent_2021}
Kaiqing Zhang, Zhuoran Yang, and Tamer Başar.
\newblock Multi-{Agent} {Reinforcement} {Learning}: {A} {Selective} {Overview} of {Theories} and {Algorithms}.
\newblock In Kyriakos~G. Vamvoudakis, Yan Wan, Frank~L. Lewis, and Derya Cansever, editors, \emph{Handbook of {Reinforcement} {Learning} and {Control}}, Studies in {Systems}, {Decision} and {Control}, pages 321--384. Springer International Publishing, Cham, 2021.
\newblock ISBN 978-3-030-60990-0.
\newblock \doi{10.1007/978-3-030-60990-0_12}.
\newblock URL \url{https://doi.org/10.1007/978-3-030-60990-0_12}.

\end{thebibliography}

\appendix

\section{DRL Hyperparameters Used}

\begin{table}[h]
\caption{Hyperparameters Used for DRL Agent Training}\label{hyperparameter_table}
\vspace{2mm}

\centering
\begin{tabular}{@{}llll@{}}
\toprule
Parameter & Value\\
\midrule
Learning Rate & 5e-5 \\
Clipping Parameter & 0.3 \\
Discount Factor ($\lambda$) & 1.0   \\
Number of SGD Iterations & 30  \\
SGD Minibatch Size & 128 \\
\bottomrule
\vspace{0mm}
\end{tabular}
\label{tab:parameters}
\end{table}

\section{Code}

Code is available on GitHub upon request.
\end{document}